\documentclass{aastex}%

\def\spose#1{\hbox to 0pt{#1\hss}} 
\def\simlt{\mathrel{\spose{\lower 3pt\hbox{$\mathchar"218$}} 
\raise 2.0pt\hbox{$\mathchar"13C$}}} 
\def\simgt{\mathrel{\spose{\lower 3pt\hbox{$\mathchar"218$}} 
\raise 2.0pt\hbox{$\mathchar"13E$}}} 
  
\def\etal{{\rm et~al.~}} 

\begin{document}

\centerline{\bf \large Constraining a possible variation of G with Type Ia supernovae}

\centerline{\large Jeremy Mould \& Syed A. Uddin\footnote{CAASTRO, http://www.caastro.org}}

\centerline{\large Centre for Astrophysics \& Supercomputing, Swinburne University}%



\begin{abstract}
Astrophysical cosmology constrains the variation of Newton's Constant
in a manner complementary to laboratory experiments, such as the celebrated
lunar laser ranging campaign. 
Supernova cosmology is an example of the former and has attained campaign
status, following planning by a Dark Energy Task Force in 2005.
In this paper we employ the full SNIa dataset to the end of 2013 to set
a limit on G variation.
In our approach we adopt the standard candle delineation of the redshift distance relation.
We set an upper limit on its rate of change   $|\dot{G}/G|$ of
0.1 parts per billion per year over 9 Gyrs. By contrast lunar laser ranging
tests variation of G over the last few decades. 
Conversely, one may adopt the laboratory result as a prior and constrain the
effect of variable G in dark energy equation of state experiments to $\delta$w~$<$~0.02. 
We also examine the parameterization G $\sim$ 1+z. Its short expansion age
conflicts with the measured values of the expansion rate and the density in a flat Universe.
In conclusion, supernova cosmology complements other experiments in limiting
G variation. An important caveat is that it rests on the assumption that the same mass of $^{56}$Ni
is burned to create the standard candle regardless of redshift.
These two quantities, f and G, where f is the Chandrasekhar mass fraction burned,
are degenerate. Constraining f variation alone requires more understanding of the SNIa mechanism.
\end{abstract}
%
\centerline{\it stars: (supernovae):general -- gravity -- (cosmology): distance scale -- 
stars: (white dwarfs)}

%

\section{INTRODUCTION }
\label{sec:intro}

The Planck mass, $m_P = \surd \hbar c/G$, is a fundamental quantity of stellar astrophysics. The Chandrasekhar mass to order unity is $m_C ~=~m_P^3/m_p^2$, where $m_p$
is the proton mass. The maximum mass of a star is approximately 
the Chandrasekhar mass multiplied by the square of the ratio of radiation pressure to gas pressure (Eddington 1917). The minimum mass of a black hole is within order
unity of the Planck mass.
Variation of Newton's constant affects
supernova cosmology via change in the Planck mass over cosmic time and
was first considered by Gazta\~{n}aga \etal (2002).
At that time there were 42 SNe available; there are now 581 (Suzuki \etal 2012).

Uzan (2003) and Narimani, Moss, \& Scott (2010, 2012) advise that constraining the constancy of dimensional
quantities is perilous. Preferred quantities are, for example, the ``gravitational
fine structure constant",  $\alpha_{\rm g}\equiv G m_{\rm p}^2/\hbar c$. 
The lunar laser ranging experiment initiated by NASA's Apollo mission is an $\alpha_{\rm g}$ experiment,
monitoring the specific potential energy of the Earth-Moon system $Gm_{\rm p}^2/ct$, where t is the time of 
flight of Earth launched photons.
Measuring the luminosity distance of galaxies, $D_L$, from type Ia supernovae  is an $\alpha_{\rm g}$ experiment, as $D_L^2 ~\propto~ m_C/m_p$, assuming a fixed fraction, f,
of $m_C$ is turned into energy and stellar luminosities are calibrated
by hydrogen burning stars. Specifically, to within a numerical constant of order unity, $m_C/m_p~=~\alpha_{\rm g}^{-3/2}$.  

Speculation about varying G began with Dirac (1937), who noted that the ratio of the electrostatic and gravitational forces between an electron and a proton
was of the same order as the number of times an electron orbits a proton
in the age of the Universe. He conjectured that $\alpha_{\rm g}$ might decay as
the inverse of cosmic time. This 20th century gravity problem (which is sometimes tackled anthropically)
has been totally eclipsed in the last decade by the cosmological constant problem (Sol\`{a} 2013). 
The contribution to the vacuum energy density of fluctuations in the gravitational field 
is larger than is observationally allowed by some 120 orders of magnitude.
Instead, the vacuum energy density is of the same order of magnitude as the present mass density of the universe. 
Although ongoing type Ia supernova observations indicate that the equation of state of Einstein's General Relativity is the best fit,
this gross cosmological constant problem provides no comfort for constant G  orthodoxy.
 
Garcia-Berro \etal (2007) review astronomical measurements and constraints on the variability of fundamental constants generally.
Garcia-Berro \etal (2006) fit a polynomial  G(z) = $G_o(1-0.01z+0.3z^2-0.17z^3)$ to the SNIa data, suggestive of a G larger in the past. 
Verbiest \etal (2008) measure orbital
period rates of pulsars and set a limit of $|\dot{G}/G| = 23 \times 10^{-12} yr^{-1}$. 
From white dwarf cooling, Garcia-Berro et al. (2011) derive an upper bound  $\dot{G}/G = -1.8 \times 10^{-12} yr^{-1}$ 
and Corsico \etal (2013) find a white dwarf pulsation limit of $\dot{G}/G = -1.3 \times 10^{-10} yr^{-1}$. 
Tomaschitz (2010) considers a  gravitational constant scaling linearly with the Hubble parameter, and fits the SNIa Hubble diagram and AGN source counts,
concluding that further observational constraints are required.

Furthermore, the luminosity of degenerate carbon core supernovae is proportional to the mass of carbon burned to $^{56}$Ni. 
The precise mechanism which powers a type Ia supernova explosion is a matter of lively debate,
 and we do not know yet whether a detonation or a deflagration occurs. What it is clear is the close correlation 
between the mass of nickel synthesised in the outburst and the luminosity.
This is discussed in quantitative experiments by Gazta\~{n}aga \etal (2002). 
The type SNIa standard candle is thought to result from a high degree of regulation
of this quantity, such as would be provided by approach to a physical limit, the Chandrasekhar mass. 
However, one may conjecture that the fraction of $m_C$ that is turned into energy may also vary with
z. Like variation of G with z, this is also an issue for constraints on the equation of state of the Universe
arising from supernova measurements.

\section{VARYING G}
There are two constraints on varying G, that which has been established
from the lunar distance since 1969 (current epoch in Table 1) and astrophysical
constraints acting over cosmic time, such as the ages of the oldest stars.
According to the theories of G variation reviewed by Faulkner (1976),
G may have been larger in the past and may be considered to follow a $t^{-1}$ decline to the current epoch. 
Williams and Dickey (2002) placed a 1$\sigma$ limit of $\dot{G}/G$ = 1.1 $\times$ 10$^{-12}$ per year 
 in recent time.
If G exceeded the present value by --3 $>$ $\dot{G}/G$ $>$ +7.3 $\times$ 10$^{-11}$ per year        
 13.7 Gyrs ago, 
and we assume the supernova luminosity scales with the Chandrasekhar mass,
we obtain Figure~1. This includes variation of the density term in the Friedmann equation, i.e. $\Omega_m$(t) 
with two cases (1)~$\Omega_m$(t) + $\Omega_\Lambda$(t) = 1 to retain flatness and (2) the dark energy density $\Omega_\Lambda$ = 0.73 with
$\Omega_k ~=~ 1~-~ \Omega_m(t) ~-~ \Omega_\Lambda(t)$. We characterize
type (2) models by the value of the curvature, $\Omega_k$, at z = 0.5.
We adopt the WMAP9 limits on curvature (Hinshaw \etal 2013), --0.0065 $<~\Omega_k~<$ 0.0012. 

\begin{figure*}
\begin{center}
\includegraphics[clip, angle=-90, width=0.99\textwidth]{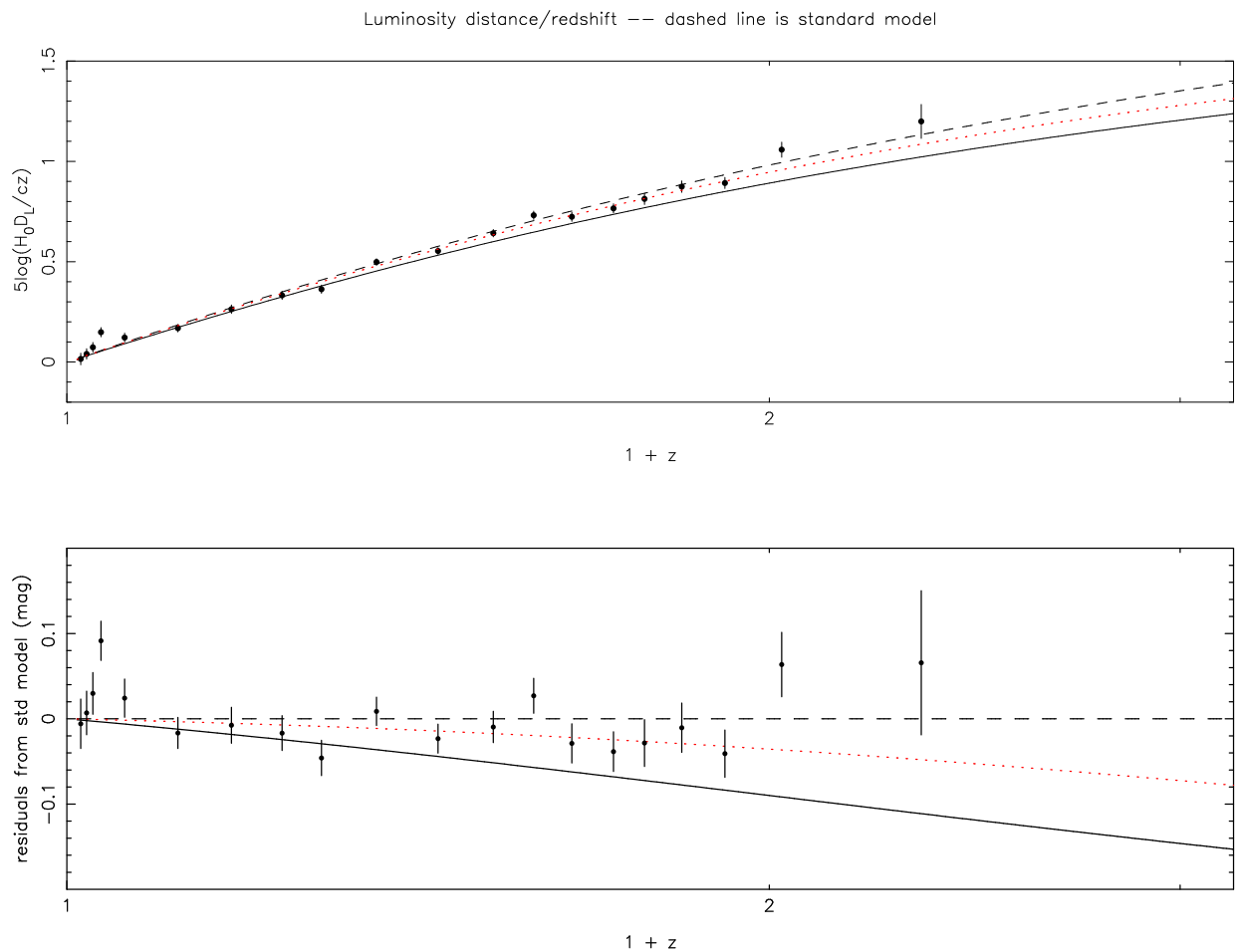}

\caption{ Luminosity distance versus redshift. The lower plot
shows residuals from the standard model.
The solid symbols are supernovae. The standard model is the dashed curve. 
The solid line and
the red line dotted line are the G varied expectations,
~the former with nonzero curvature. The most distant SNIa is at 9 billion light years in the standard cosmology with H$_0$ = 73.8 $\pm$ 2.4 km/s/Mpc,
given by Riess \etal (2012).}
\end{center}
\end{figure*}

\subsection{The Supernova Ia Constraint}
The current supernova data (Suzuki \etal 2012) are shown in Figure 1.
If we assume $\Lambda$CDM cosmology with w = --1, current SNIa data constrain
G stability to $\dot{G}/G$ = (--3,+7.3) $\times$ 10$^{-11}$ per year,
This constraint is obtained if we adopt the standard model of cosmology as a prior.
We now (1) reexpress this as a constraint on  $\dot{\alpha_{\rm g}}/\alpha_{\rm g}$,
and (2) invert the argument to constrain $w$, given laboratory limits on G variation.

\begin{figure*}
\begin{center}
\includegraphics[angle=-90, width=0.5\textwidth]{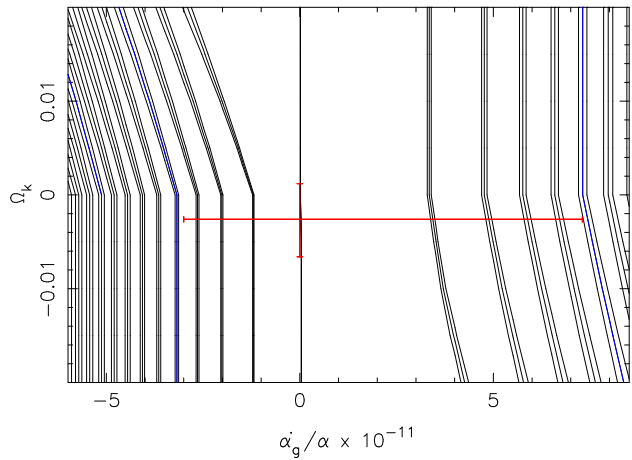}
\caption{ A vertical central valley of parameter space is permitted by these
$\chi^2$ contours. The first contour on either side of zero is $\chi^2$ per degree of freedom = 1.2 
with a spacing between triplet contours of 0.2. 
Positive values of $\dot{\alpha_{\rm g}}$ 
have the sense of G larger in the past.
With the WMAP9 limits on curvature (Hinshaw \etal 2013)
shown by the vertical error bar this implies
a SNIa cosmology constraint on G stability  in the standard
cosmology of (--3,+7.3) $\times$ 10$^{-11}$ per year, evaluated at $\chi^2$ = 2.}
\end{center}
\end{figure*}

(1) To determine the upper limit on $\dot{\alpha_{\rm g}}/\alpha_{\rm g}$
~we calculate $\chi^2$ to compare the data with the prediction, marginalizing over H$_0$, and show this in Figure 2.
The contours of $\chi^2$ are oriented close to vertical, resulting in clear limits on G variation.
 This constraint, our main result, --3 $<~10^{11}$ 
$\dot{G}/G$ 
$<$ 7.3~ per year, may be expected
to strengthen towards parity with the laboratory, in the era of dark energy experiments such as LSST,
see Weinberg \etal (2013).
An equivalent dimensionless limit is  --0.5 $<~\dot{G}/(GH_0)~<$  1, where 1/$H_0$ is the age of the Universe.

\begin{table*}
\centering
\caption{\bf : Constraints on the rate of variation of the gravitational constant.}
\begin{tabular}{@{}cllcll@{}}
\hline\hline
\vspace{1mm}
\\ 
$\dot{G}/G$  
&  Current epoch & &$\dot{G}/G$  
&  Cosmic time\\
$10^{-13}~yr^{-1}$ &&   &  $10^{-13}~yr^{-1}$ &  \\
\hline%
2$\pm$7  & lunar laser ranging& (1) &             0  $\pm$4 &  big bang nucleosynthesis& (2)\\
40 $\pm$50 &  binary pulsar &(3) &                -1.42+2.48-2.27 &    Planck+WMAP+BAO &(4)\\                   
230        & PSR J0437--4715 &(7)        &              0$\pm$16  & helioseismology& (5)\\
 &  &&   -6$\pm$20  & neutron star mass &(6)\\
&&&--300,+730&this paper\\
\hline\hline
\end{tabular}
\begin{center}
\end{center}
\begin{tabular}{@{}l@{}}

$Notes$ : the uncertainties are 1$\sigma$  unless otherwise noted. \\
1: Muller \& Biskupek 2007; 2:Copi \etal  2004; 3: Kaspi \etal 1994\\
4: Li \etal  2013; 5: Guenther \etal 1998; 6: Thorsett 1996 7: Verbiest \etal 2008\\
\end{tabular}
\end{table*}

(2) This constraint is obtained if we adopt the standard model of cosmology as a prior.
However, these SNIa data are conventionally used as a measurement of $\Omega_\Lambda$.
There is therefore a degeneracy between this and $\dot{G}$ addressed by the same data.
We can quantify the degeneracy using the generalization of the Friedmann equation as a polynomial by Mould (2011).
Mould showed that if, such a polynomial is adopted to fit Figure 1, 
$$(H/H_0)^2 = \Sigma_n~(1+z)^n~\Omega_n = h^2(z),~~~~~~~~~~\eqno(1)$$
\noindent relationships (degeneracies) between the $\Omega_n$ coefficients result from the limited available constraints (SNIa, CMB).
If the SNIa data are  used to constrain the equation of state of the Universe
with w~$\approx$~--1, there is therefore a degeneracy between w and $\dot{G}$ addressed by the same data.
For z $\sim$ 1 and zero curvature, $\delta w \approx 2 \delta \Omega_m$ from equations (3) \& (11) of Mould (2011).
For G stability to  2 parts in 10$^{12}$ per year, $\delta\alpha_{\rm g}/\alpha_{\rm g}$
= 0.0137 at z = 1,
which corresponds to $\delta w \approx$ 0.03. The current experimental uncertainty in w (Rapetti \etal 2013) is 0.07.
Both quantities therefore need to be constrained jointly.
On the other hand, if one is prepared to adopt the lunar laser ranging
results as a prior on dark energy experiments valid over all of cosmic time, 
the effect of variable G is constrained so that $\delta$w $<$ 0.02 (95\% confidence).
This is not a negligible contribution to the $w$ error budget,
 and  it should not be ignored ($cf.$ 
	Mortonson \etal 2014).

Finally, the coupling of $D_L$, f, and $m_C$ is direct. Analytically, 2$\delta D_L/D_L$ = $\delta$f/f = --1.5$\delta$G/G = $\delta m_C/m_C$.  
Our limit on $\delta\alpha_{\rm g}/\alpha_{\rm g}$ 
~is thus degenerate with an equivalent limit on $\delta$f/f. 

\subsection{Other Parameterizations}
Pragmatically, the key result here devolves from an assumed t$^{-1}$ variation of  $\alpha_{\rm g}$.
Its basis is historical and traces back to Dirac's (1937) large numbers hypothesis and the steady state Universe,
neither of which have any real traction today.
Other parameterizations are possible and even natural, such as 1+z scaling. 
One form is $\alpha_{\rm g}$ = $\alpha_0$ + $\alpha^\prime z$.
In this case we obtain --0.02~$<~\alpha^\prime/\alpha_0~<$~0.04.

If G $\sim$ 1+z, the $\Omega_3$ coefficient in equation (1) is promoted to $\Omega_4$; that is, it becomes an anti-radiation pressure term.
Assuming $\Omega_1$ = 0, the resulting degeneracies can be expressed~(Mould 2011)
$$(f_0-f_2)\delta\Omega_0 = (f_2-f_4)\delta\Omega_4 ~~~~~~~~~~\eqno(2)$$
where 
$$f_n = \int_0^z(1+z^\prime)^nh^{-3}(z^\prime)dz^\prime .$$

Any G variation that scales as 1+z is traded off against  $\Omega_\Lambda$,
according to (from Table 2) $\delta\Omega_\Lambda ~=~ \delta\Omega_0 ~=~ 301.5 \delta\Omega_4$,
when SNIa and CMB anisotropy data measure cosmological parameters simultaneously.
 A universe with just conventional dark energy and  ``radiation" like this has an age in units of 1/H$_0$ 
obtained by integrating equation (1) with $a^{-1}$ = 1 + z.

$$t = \int_0^1 \frac{a da}{\dot{a} a} = \int_0^1 \frac{da}{H_0 a \surd(\Sigma_n \Omega_n a^{-n})}~~~~\eqno(3)$$.
$$tH_0 = \int_0^1 \frac{da}{a\surd(\Omega_0+\Omega_m a^{-4})} = \int_b^\infty \frac{dy}{2\surd \Omega_0 y\surd(1+y^2)} $$
$$= \frac{1}{2\surd\Omega_0}\int {\rm cosech(x)} dx 
= \frac{1}{2\surd\Omega_0}[\rm{ln}|\rm{tanh}(x/2)|]_{arsinh(b)}^\infty$$
$$ = \frac{1}{2\surd\Omega_0}\rm{ln}~(\rm{tanh (arsinh}(b)/2))~~~~~~~~~\eqno(4)$$ 

where $$b^2 ~=~ \Omega_m/\Omega_0 ~=~ \Omega_m/(1-\Omega_m) $$
and $$y = ba^{-2} = {\rm sinh}(x)~~~~~~~~~~~\eqno(5)$$. 

For $ \Omega_M$ = 0.27, x = 0.575 at z = 0 and the age is  0.745.
This is a second contradiction with the standard model of cosmology, as Planck finds an age approximately one in these units (Ade et al 2013, Efstathiou 2013).
A further contradiction with the age of the globular clusters is mildly ameliorated by higher central temperatures of stars (GM$m_p/kR$, where M, R are the stellar mass and radius) during the epoch of reionization, when they were formed, and the extraordinary temperature sensitivity of the CN cycle of fusion, but for most of the low mass stars' lifetime core temperatures are close to normal and ages are only mildly affected (Vandenberg 1977).

How severe a constraint on G $\sim$ 1+z is this? 
Error analysis gives terms in $\delta\Omega_M/\Omega_M , \delta\Omega_\Lambda / \Omega_\Lambda {\rm ~and~ } \delta H_0/H_0$. 
The first and last of these are of order a few percent and the second is smaller. 
This parameterization can therefore be rejected with 99\% confidence. 
G~$\sim~(1~+~z)^{1/n}$ would be less unacceptable for large n, but is not a natural parameterization.

\pagebreak
\vskip 0.25 truein

\leftline{\bf Table 2: Equation of state components}

\begin{tabbing}

ssssssssssssssss\=sssssssssss\=ssssssssssss\=sssssssssssss\=sssssssssssss\=sssss\kill
$\Omega_n$\>n\>w$_n$\>f$_n$\\
vacuum\>0\>--1\>0.662\\
textures\>1\>--2/3\>0.964\\
``curvature''\>2\>--1/3\>2.294\\
matter\>3\>0\>10.40\\
radiation\>4\>+1/3\>494.4\\
\\
The $f_n$ coefficients have been evaluated at $\Omega_3$ = 0.27.
\end{tabbing}

\section{SUMMARY}
Our conclusions from this work are as follows.
\begin{enumerate}
\item The validity of the SNIa standard candle depends on the stability of G and the stability of f,
the fraction of the Chandrasekhar mass turned into energy. We have considered the former in this paper
and derived a constraint on the gravitational fine structure constant which can be compared
with other combined astrophysical-cosmological constraints. But this is inextricably degenerate
with possible evolution of f due to changes over cosmic time of SNIa progenitor astrophysics.
With this caveat we set a SNIa cosmology constraint on G stability  in the standard
cosmology of (--3,+7.3) $\times$ 10$^{-11}$ per year.

\item This limit is two orders of magnitude weaker than that from lunar laser ranging.
But that is a current epoch result and complements, but does not replace, a constraint that spans cosmic time.

\item The limit is also two orders of magnitude weaker than that arising from the great sensitivity
to density of Big Bang nucleosynthesis. The SNIa standard candle, however, has the distinct advantage
of covering the last 10$^{10}$ years of cosmic time, rather than the first twenty minutes.

\item Our result is an update of Gazta\~{n}aga \etal (2002), who found $\dot{G}/G~<$
12 $\times$ 10$^{-12}~ h_{70}$ /yr for $\Omega_\Lambda$ = 0.8, $\Omega_M$ = 0.2.
This is a 1$\sigma$ limit, like ours, and directly comparable since $h_{70}~\approx$ 1.
The order of magnitude more supernovae now available have allowed us to relax their flat Universe assumption,
but has also relaxed their limit on $\dot{G}$. 

\item Caution would dictate that experiments to measure the equation of state of the Universe carry the caveat
that f and G stability is assumed. For the latter, lunar laser ranging is available as a prior
and limits $\delta w$ to 0.02, but the former has not been quantified and demands further understanding
of the SNIa mechanism.

\item A parameterization G $\sim$ 1+z is interesting on theoretical grounds related to the unity of forces. 
However, with such an equation of state the expansion age of the Universe is too short.
That parameterization can therefore be rejected with 99\% confidence.

\end{enumerate}

\acknowledgements
We are grateful to Chris Blake and Michael Murphy for helpful advice and to an anonymous referee
for emphasizing the issue of the quantum of $^{56}$Ni.
CAASTRO is the ARC's Centre of Excellence for All-Sky Astrophysics, funded by grant CE11001020.


\section*{References}
\noindent Ade, P. et al 2013, A\&A, in press, astro-ph 1303.5076\\
Copi, C.  Davis, A. \& Krauss, L. 2004, Phys. Rev.Lett. 92, 171301\\
Corsico, A., \etal 2013, JCAP, 06, 032\\
Dirac, P. 1937, Nature, 139, 323\\
Eddington, A. S. 1917, MNRAS, 77, 596\\
Efstathiou, G. 2013, astro-ph 1311.3461\\
Faulkner, D. 1976, MNRAS, 176, 621\\
Garcia-Berro, E. \etal 2006, IJMPD, 15, 1163\\
Garcia-Berro, E. \etal 2007,  A\&ARv, 14, 113\\
Garcia-Berro, E. \etal 2011,  JCAP, 05, 21\\
 Gazta\~{n}aga, E. \etal 2002, Phys Rev D, 65, 023506\\
Guenther, B., Krauss, L. \& Demarque, P. 1998, ApJ 498, 871\\ 
Hinshaw, G. \etal 2013, ApJS, 208, 19\\
Kaspi, V., Taylor, J. \& Ryba, M. 1994, ApJ 428, 713\\ 
Li, Y.-C \etal 2013, Phys Rev D, 88, 084053\\
Mortonson, M. \etal 	 2014, arXiv, 1401.0046\\	
Mould, J. 2011, PASP, 123, 1030\\
Muller, J. \& Biskupek, L. 2007, Class. Quant. Grav. 24, 4533\\ 
Narimani, A. \etal 2012, Ap\&SS, 341, 617\\
Narimani, A. \etal 2010, IJMPD, 19, 2289\\
Rapetti, D. \etal 2013, MNRAS, 432, 973\\
Riess, A. \etal 2012, ApJ, 752, 76\\
Sol\`{a}, J. 2013, JPhCS, 453, 2015\\
Suzuki, N., \etal 2012, ApJ, 746, 85\\
Thorsett, S. 1996, Phys. Rev. Lett. 77, 1432 \\
Tomaschitz, R.	2010, Ap\&SS, 325, 259\\
Uzan, J.-P. 2003, RvMP, 75, 403\\
Vandenberg, D. 	1977, MNRAS, 181, 695\\
Verbiest, J. \etal	2008, ApJ, 679, 675\\
Weinberg, D. \etal 2013, PhR 530, 87\\
Williams,~J. \& Dickey,~J. 2002, http://ilrs.gsfc.nasa.gov/docs/williams\_lw13.pdf



\end{document}